\documentclass[12pt]{article}

\usepackage{color}
\definecolor{darkblue}{rgb}{0.1,0.1,.7}
\usepackage[colorlinks, linkcolor=darkblue, citecolor=darkblue, urlcolor=darkblue, linktocpage]{hyperref}
\usepackage[]{amsmath}
\usepackage[]{graphicx}
\usepackage[]{latexsym}
\usepackage{slashed,graphicx,color,amsmath,amssymb}
\usepackage[mathscr]{eucal}
\usepackage{mathrsfs}
\usepackage{geometry}
\usepackage[margin=10pt,font=small,labelfont=bf]{caption}
\usepackage{amscd}
\usepackage{bm}
\usepackage{xcolor}
\usepackage{upgreek}
\usepackage[square, comma, sort&compress,numbers]{natbib}
\usepackage[all,cmtip]{xy}
\numberwithin{equation}{section}
\newcommand{\rg}{\mathrm{g}}

\newcommand{\rE}{\mathrm{E}}

\newcommand{\rH}{\mathrm{H}}

\newcommand{\rM}{\mathrm{M}}

\newcommand{\rP}{\mathrm{P}}

\newcommand{\rT}{\mathrm{T}}

\begin{document}
\vspace*{-.6in} \thispagestyle{empty}
\begin{flushright}
{CERN-TH-2019-030}
\end{flushright}
\vspace{.2in} {\Large
\begin{center}
{\bf Expectation value of $\rT\overline{\rT}$ operator in curved spacetimes}
\end{center}}
\vspace{.2in}
\begin{center}
Yunfeng Jiang
\\
\vspace{.3in}
\small{\textit{Theoretical Physics Department, CERN, Geneva, Switzerland}
}

\end{center}

\vspace{.3in}

\begin{abstract}
\normalsize{We study the expectation value of the $\rT\overline{\rT}$ operator in maximally symmetric spacetimes. We define an diffeomorphism invariant biscalar whose coinciding limit gives the expectation value of the $\rT\overline{\rT}$ operator. We show that this biscalar is a constant in flat spacetime, which reproduces Zamolodchikov's result in 2004. For spacetimes with non-zero curvature, we show that this is no longer true and the expectation value of the $\rT\overline{\rT}$ operator depends on both the one- and two-point functions of the stress-energy tensor.}
\end{abstract}

\vskip 1cm \hspace{0.7cm}

\newpage

\setcounter{page}{1}
\begingroup
\hypersetup{linkcolor=black}
\tableofcontents
\endgroup

%%%%%%%%%%%%%%%%%%%%%%%%%%%%%%%%%%%%%%%%%%%%%%%%%%%%%%%%%%%%%%%%%%%%%%%%
\section{Introduction}
\label{sec:1}
%%%%%%%%%%%%%%%%%%%%%%%%%%%%%%%%%%%%%%%%%%%%%%%%%%%%%%%%%%%%%%%%%%%%%%%%
The $\rT\overline{\rT}$ deformation \cite{Smirnov:2016lqw,Cavaglia:2016oda} is a special kind of irrelevant deformation of 2d quantum field theory triggered by the composite operator $\rT\overline{\rT}$ \cite{Zamolodchikov:2004ce}. Unlike usual irrelevant deformations, which are typically ambiguous and complicated, the $\rT\overline{\rT}$ deformation is exactly solvable. More precisely, the deformed finite volume spectrum \cite{Smirnov:2016lqw,Cavaglia:2016oda}, torus partition function \cite{Datta:2018thy,Cardy:2018sdv,Aharony:2018bad,Dubovsky:2017cnj,Dubovsky:2018bmo} and the deformed $S$-matrix \cite{Dubovsky:2013ira,Dubovsky:2017cnj} can be determined explicitly. The $\rT\overline{\rT}$ deformed theory provides a novel type of UV behavior for QFT called asymptotic fragility which was first proposed in \cite{Dubovsky:2012wk}. In this case, the UV theory is not a fixed point since it is not a local quantum field theory. Nevertheless, the non-local UV theory is perfectly well-defined since many interesting physical observables can be computed explicitly.\par

Intriguingly, the non-locality and solvability of this deformation can be understood from a more geometrical perspective. An infinitesimal $\rT\overline{\rT}$ deformation can be interpreted as summing over variations of the underlying spacetime geometry \cite{Cardy:2018sdv}. A full path integral definition for the deformed theory is provided by coupling the QFT to Jakiw-Teitelboim (JT) gravity\footnote{More precisely, a slightly modified version of JT gravity.} \cite{Dubovsky:2017cnj,Dubovsky:2018bmo}. The later can in turn be understood as a dynamical change of coordinates (or field dependent change of coordinates) \cite{Conti:2018tca,Conti:2018jho}, at least at the classical level.\par

What's more, the torus partition function of the $\rT\overline{\rT}$ deformed conformal field theory (CFT) is still modular invariant \cite{Datta:2018thy}, although the deformed theory is neither conformal nor local. More surprisingly, by requiring modular invariance and that the spectrum is deformed in a universal way, one can fix the deformed theory uniquely to be the $\rT\overline{\rT}$ deformed CFT \cite{Aharony:2018bad}. From modularity of the torus partition function, one can derive the asymptotic density of states of the $\rT\overline{\rT}$ deformed CFT\footnote{For the sign of the deformation parameter where the deformed spectrum is real for high energy states.} and find that it interpolate between the Cardy behavior and the Hagedorn behavior from IR to UV. This fact makes the $\rT\overline{\rT}$ deformed CFT a promising candidate for holographic dual for a large class of vacua of string theory
in asymptotically flat linear dilaton spacetimes \cite{Giveon:2017nie}.\par

In the holographic side, for one sign of the deformation parameter, the holographic dual corresponds to a cut-off geometry in the bulk \cite{McGough:2016lol,Kraus:2018xrn}. This is generalized to higher dimensions \cite{Hartman:2018tkw,Taylor:2018xcy} and to sphere partition function \cite{Caputa:2019pam}. Very recently, it was pointed out \cite{Guica:2019nzm} that the cut-off geometry interpretation is only valid for the pure gravity sector in the bulk. In the general case, it is more appropriate to interpret the holographic dual in terms of a mixed boundary condition. Parallel to the development in QFT, a single trace deformation for the string worldsheet model has been proposed \cite{Giveon:2017myj,Giveon:2017nie,Asrat:2017tzd,Chakraborty:2018kpr}. For other related interesting developments, we refer to \cite{Shyam:2017znq,Giribet:2017imm,Cottrell:2018skz,Aharony:2018vux,Bonelli:2018kik,Baggio:2018gct,Baggio:2018rpv,
Chang:2018dge,Chen:2018eqk,Araujo:2018rho,Sun:2019ijq,Park:2018snf,Wang:2018jva,Santilli:2018xux,
Chakraborty:2018aji,Shyam:2018sro,Babaro:2018cmq,Akhavan:2018wla,Hashemi:2019xeq}

For theories with an additional $U(1)$ current, one can define a similar solvable deformation called $J\bar{T}$ deformation both in quantum field theory \cite{Guica:2017lia,Bzowski:2018pcy,Guica:2019vnb,Aharony:2018ics,Nakayama:2018ujt} and analogously on the string worldsheet \cite{Chakraborty:2018vja,Apolo:2018qpq}.

It is interesting to see whether some of these nice features can be generalized to curved spacetime. We have at least two motivations. The first comes from the JT gravity interpretation of the $\rT\overline{\rT}$ deformation in flat spacetime. It is interesting to see whether this interpretation is still true in curved background, in particular in $AdS_2$. There has been many exciting progress in understanding JT quantum gravity on $AdS_2$ \cite{Maldacena:2016upp,Yang:2018gdb,Kitaev:2017awl,Kitaev:2018wpr}. If JT gravity has an alternative interpretation as $\rT\overline{\rT}$ deformation of usual QFT. This will shed new lights on both subjects. The first step towards testing such a relationship is thus defining the $\rT\overline{\rT}$ deformation for QFT on curved background. The second motivation is to understand more general solvable deformations, such as the one related to the dS/dS correspondence proposed in \cite{Gorbenko:2018oov}.

One of the central reasons underlying the solvability of the $\rT\overline{\rT}$ deformation is the factorization formula for the expectation value of the $\rT\overline{\rT}$ operator. This was first proved by Zamolodchikov \cite{Zamolodchikov:2004ce} (see also \cite{Cardy:2018sdv} for a slightly different proof). He argued that the composite operator $\rT\overline{\rT}$ is well-defined up to total derivatives using point splitting. Then he proved that the expectation value of this composite operator can be computed exactly in terms of the expectation values of the stress energy tensor
\begin{align}
\label{eq:fac}
\langle n|\rT\overline{\rT}|n\rangle=\langle n|T|n\rangle\langle n|\bar{T}|n\rangle-\langle n|\Theta|n\rangle^2.
\end{align}
We will denote this relation by
\begin{align}
\langle\rT\overline{\rT}\rangle=\langle\rT\rangle\langle\overline{\rT}\rangle.
\end{align}
Therefore, the first step towards $\rT\overline{\rT}$ deformation in curved spacetime is studying the expectation value of the $\rT\overline{\rT}$ operator. In particular, we want to see whether (\ref{eq:fac}) still holds in the presence of non-zero curvature.
There has been some works which involve $\rT\overline{\rT}$ deformation in curved spacetime. However, all these works assume large $c$ limit where large $c$ factorization guarantees the factorization formula \cite{Donnelly:2018bef,Hartman:2018tkw,Caputa:2019pam}.
In this paper, we fill this gap and analyse the expectation value of $\rT\overline{\rT}$ more carefully. We show that the factorization does not hold at finite $c$ in curved spacetime. Let us make one comment on the states in which we compute the expectation value. In Zamolodchikov's original work \cite{Zamolodchikov:2004ce}, the state $|n\rangle$ is any energy-momentum eigenstate on a cylinder. In what follows, we consider the expectation values of the fields in the maximally symmetric states $|\psi\rangle$. We will denote $\langle\mathcal{O}_1(x_1)\cdots\mathcal{O}_n(x_n)\rangle\equiv\langle\psi|\mathcal{O}_1(x_1)\cdots\mathcal{O}_n(x_n)|\psi\rangle$. The requirement of maximal symmetry is necessary to insure the decomposition (\ref{eq:ansatzTT}) which is crucial for our derivation.\par

In this work, we focus on the curved spacetime with maximal symmetry and constant curvature. The curvatures for these spaces are simply given by
\begin{align}
\mathcal{R}_{\mu\nu\rho\sigma}=\pm\frac{1}{R^2}\left(g_{\mu\sigma}g_{\nu\rho}-g_{\mu\rho}g_{\nu\sigma}\right),\qquad
\mathcal{R}=\pm\frac{d(d-1)}{R^2}
\end{align}
where $R$ is the scale and $d$ is the dimension of spacetime. The non-zero curvature can be either positive or negative. The positive curvature spaces include sphere and de Sitter space while the negative curvature spaces include Poincar\'e disc and anti-de Sitter space. Our main result is the following formula for the expectation value of $\langle\rT\overline{\rT}\rangle$ in the maximally symmetric spacetimes
\begin{align}
\label{eq:mainresult}
\langle\rT\overline{\rT}\rangle=\langle\rT\rangle\langle\overline{\rT}\rangle
-\int_0^{\theta_{\text{max}}}\rg(\theta)\mathsf{\Delta}_{\text{con}}(\theta)d\theta
\end{align}
where
\begin{align}
\label{eq:deltaLR}
\mathsf{\Delta}_{\text{con}}(\theta)=\left(g_{\mu\nu}-n_{\mu}n_{\nu}\right)g_{\alpha'\beta'}
\left[\langle T^{\mu\nu}(x)T^{\alpha'\beta'}(y)\rangle-\langle T^{\mu\nu}(x)\rangle\langle T^{\alpha'\beta'}(y)\rangle\right]
\end{align}
Equivalently, (\ref{eq:mainresult}) can be written as a line integral
\begin{align}
\langle\rT\overline{\rT}\rangle=\langle\rT\rangle\langle\overline{\rT}\rangle
-\int_{\gamma_{\max}}\rg(\theta(x,y))\mathsf{\Delta}_{\text{con}}(\theta(x,y))n_{\mu}dx^{\mu}
\end{align}
where $\gamma_{\max}$ is the geodesic between $y$ and $y_{\text{max}}$ such that $\theta(y,y_{\text{max}})=\theta_{\text{max}}$.
Here $\theta(x,y)$ is the geodesic distance between the two spacetime points $x$ and $y$ and $n_{\mu}=\partial_\mu\theta(x,y)$ where the derivative acts on point $x$. It can be shown that the rhs of (\ref{eq:deltaLR}) is a function only depends on $\theta$. $\rg(\theta)$ is a known function given by
\begin{align}
&\mathcal{R}>0:\qquad \rg(\theta)=-\frac{2}{R}\left(\sin\frac{\theta}{2R}\right)\left(\cos\frac{\theta}{2R}\right)^3,\\\nonumber
&\mathcal{R}<0:\qquad \rg(\theta)=\frac{2}{R}\left(\sinh\frac{\theta}{2R}\right)\left(\cosh\frac{\theta}{2R}\right)^3.
\end{align}
The upper bound for the integral $\theta_{\text{max}}$ is the maximal value for the geodesic distance in the given space. For positive curvature spacetime, $\theta_{\text{max}}=\pi R$; for negative curvature spacetime, $\theta_{\text{max}}\to\infty$.\par

In flat spacetime ($R\to\infty$) or large $c$ limit $(c\to\infty)$, the second term on the rhs of (\ref{eq:mainresult}) vanishes and we indeed have factorized formula. However, for finite $R$ and finite $c$, the deviation from factorized result is given by the second term which depends on the information of two-point functions of the stress-energy tensor.\par

The rest of the paper is organized as follows. In section~\ref{sec:fac} we review the derivation of factorized formula in flat spacetime and define an invariant biscalar on curved spacetime which plays an important role in computing the expectation value of the $\rT\overline{\rT}$. This invariant biscalar is a specific projection of the two-point function of stress-energy tensor. In section~\ref{sec:maxTensor}, we review some basic properties and the tensor decomposition of two-point functions of stress-energy tensor in maximally symmetric spacetime. In section~\ref{sec:invariantC2d} we focus on 2d spacetime and derive a differential equation for the biscalar based on the symmetry of spacetime and the conservation of stress-energy tensor. Using this differential equation, we derive the expression for the expectation value of the $\rT\overline{\rT}$ operator in section~\ref{sec:expectV}. In section~\ref{sec:example} we consider some example using explicit coordinate systems. In section~\ref{sec:flatd}, we comment on a similar analysis in higher dimensions. We conclude in section~\ref{sec:conclude}.

%%%%%%%%%%%%%%%%%%%%%%%%%%%%%%%%%%%%%%%%%%%%%%%%%%%%%%%%%%%%%%%%%%%%%%%%%%%
\section{An invariant biscalar}
\label{sec:fac}
%%%%%%%%%%%%%%%%%%%%%%%%%%%%%%%%%%%%%%%%%%%%%%%%%%%%%%%%%%%%%%%%%%%%%%%%%%%
In this section, we define a biscalar $\mathsf{C}(x,y)$ which is a scalar that is supported at two spacetime points $x^\mu$ and $y^{\mu}$. This biscalar should be invariant under diffeomorphism. The coinciding limit $x\to y$ of this quantity should give the expectation value of the $\rT\overline{\rT}$ operator. We first recall the derivation of factorization formula in flat spacetime and then give the definition of $\mathsf{C}(x,y)$ in curved spacetime.

\paragraph{Flat spacetime derivation} We review the derivation of factorization formula following Cardy \cite{Cardy:2018sdv}. In 2d flat spacetime we define the following quantity in the Cartesian coordinate
\begin{align}
\label{eq:coord}
\mathsf{C}(x,y)=&\,\left(\delta_{ik}\delta_{jl}-\delta_{ij}\delta_{kl} \right)\langle T^{ij}(x)T^{kl}(y)\rangle
=\langle T^{ij}(x)T_{ij}(y)\rangle-\langle T^{i}_{i}(x)T^j_j(y)\rangle\\\nonumber
=&\,\epsilon_{ik}\epsilon_{jl}\langle T^{ij}(x)T^{kl}(y)\rangle
\end{align}
We use Latin letters $i,j,...$ to denote indices of Cartesian system and Greek letters $\mu,\nu,...$ for those of general coordinate systems. In order to see that $\mathsf{C}(x,y)$ is a constant, we need to prove that
\begin{align}
\label{eq:partialC}
\frac{\partial}{\partial x^m}\mathsf{C}(x,y)=0
\end{align}
In Cartesian coordinate, we have
\begin{align}
\frac{\partial}{\partial x^m}\epsilon_{ik}=\epsilon_{mk}\frac{\partial}{\partial x^i}+\epsilon_{im}\frac{\partial}{\partial x^k}
\end{align}
Therefore
\begin{align}
\frac{\partial}{\partial x^m}\mathsf{C}(x,y)=&\,\epsilon_{jl}
\left[\epsilon_{mk}\frac{\partial}{\partial x^i}+\epsilon_{im}\frac{\partial}{\partial x^k} \right]\langle T^{ij}(x)T^{kl}(y)\rangle,\\\nonumber
=&\,\epsilon_{jl}\epsilon_{im}\frac{\partial}{\partial x^k}\langle T^{ij}(x)T^{kl}(y)\rangle,\\\nonumber
=&\,-\epsilon_{jl}\epsilon_{im}\frac{\partial}{\partial y^k}\langle T^{ij}(x)T^{kl}(y)\rangle=0
\end{align}
where in the first line and third line we used the conservation of stress energy tensor $\partial_i T^{ij}=0$. More precisely, we used the resulting Ward identity of the two-point function. In the second line, we use the fact that $\langle T^{ij}(x)T^{kl}(y)\rangle$ is a function that only depends on $(x-y)^2$. This is due to translational invariance of the spacetime and we have
\begin{align}
\label{eq:xToy}
\left(\frac{\partial}{\partial x^k}+\frac{\partial}{\partial y^k}\right)\langle T^{ij}(x)T^{kl}(y)\rangle=0.
\end{align}
Since $\mathsf{C}(x,y)$ is a constant, we can take $x$ and $y$ anywhere in the spacetime. We can take the coinciding limit $y\to x$. It has been shown by Zamolodchikov \cite{Zamolodchikov:2004ce} this limit is well-defined and $\mathsf{C}(x,x)=\langle\rT\overline{\rT}\rangle$. On the other hand, one can also take the limit where $|x-y|\to\infty$. By the cluster decomposition theorem, the two-point function decomposes to the product of one-point functions. We thus arrive at the factorization formula
\begin{align}
\label{eq:facTTflat}
\langle\rT\overline{\rT}\rangle=\epsilon_{ik}\epsilon_{jl}\langle T^{ij}\rangle\langle T^{kl}\rangle.
\end{align}
This relation states that the expectation value of the composite $\rT\overline{\rT}$ operator can be expressed in terms of the expectation value of the stress-energy tensor. From the derivation, we see that the following three ingredient are important to prove the factorization formula
\begin{enumerate}
\item Introducing the quantity $\mathsf{C}(x,y)$. This quantity is a bridge between the lhs and rhs of (\ref{eq:facTTflat}).
\item The fact that $\langle T^{ij}(x) T^{kl}(y)\rangle$ only depends on $(x-y)^2$, which is due to the symmetry of the spacetime.
\item Conservation of stress-energy tensor, or equivalently, the Ward identity of the two-point function of the stress-energy tensor.
\end{enumerate}
We shall see that all the three ingredients have natural generalizations to constant curvature spacetime. In this section, we discuss the generalization of the first ingredient, namely the definition of $\mathsf{C}(x,y)$ to curved spacetime.

\paragraph{Parallel propagator} In order $\mathsf{C}(x,y)$ to be physical, we require that it is invariant under diffeomorphism. The form in flat spacetime (\ref{eq:coord}) is suggestive but somewhat misleading.
It is tempting to simply replace $i,j$ by $\mu,\nu$ in the first line of (\ref{eq:coord}) and take it as the definition for $\mathsf{C}(x,y)$ in curved spacetime. However, this naive replacement does not lead to a good definition except at the coinciding limit. The reason is that we are contracting indices at different spacetime points and the resulting quantity is not invariant under local coordinate transformations. To motivate our definition in what follows, let us consider a similar situation in gauge theory. Suppose we want to make a bilinear quantity in terms of fermions $\bar{\psi}(x)$ and $\psi(y)$ at two different spacetime points that is invariant under local gauge transformation
\begin{align}
\bar{\psi}(x)\mapsto e^{-i\alpha(x)}\bar{\psi}(x),\qquad \psi(y)\mapsto e^{i\alpha(y)}\psi(y)
\end{align}
Simply taking $\bar{\psi}(x)\psi(y)$ does not work since the two phase factors do not cancel. The solution in this case is well-known. To make a gauge invariant quantity, we need a Wilson line to connect the two spacetime points. The following quantity
\begin{align}
\bar{\psi}(x)W(x,y)\psi(y)
\end{align}
is gauge invariant. Here $W(x,y)$ is the Wilson line
\begin{align}
\label{eq:WL}
W(x,y)=\mathrm{P}\exp\left(i\int_{\gamma}A_{\mu}(x')d{x'}^{\mu} \right)
\end{align}
where $\rP$ denotes path ordering and $\gamma$ is a path connecting the two spacetime points.\par

To define an invariant quantity in our case, we also need certain kind of ``connection'' which connects two spacetime points, similar to a Wilson line in gauge theory. Such a quantity is called the parallel propagator which we denote by $I_{\mu\alpha'}(x,y)$. The parallel propagator is a bi-vector which connects two spacetime points. Interestingly, it can be written as (see for example Appendix I of \cite{Carroll:1997ar})
\begin{align}
I^{\mu}_{\phantom{\mu}\nu}(x,y)=\rP\exp\left(-\int_{\gamma}\Gamma^{\mu}_{\sigma\nu}(x')d{x'}^{\sigma}  \right)
\end{align}
where we basically replace the gauge connection in (\ref{eq:WL}) by the affine connection. The full definition of $I^{\mu}_{\phantom{\mu}\nu}(x,y)$ depends on the choice of the path $\gamma$. In what follows, we choose $\gamma$ to be the geodesic that connects the two spacetime points. When there are more than one geodesics between the two points, we choose the one with shortest distance\footnote{For the case where all the geodesics have the same distance, we can choose any one of them. This happens for example, for the two antipodal points of the sphere.}. This is because it appears naturally in the spacetimes with maximal symmetry, as we will discuss below.
For a more detailed discussion of parallel propagator, we refer to \cite{Carroll:1997ar,Allen:1985wd,Osborn:1999az}. Using the parallel propagator, we propose that the invariant biscalar $\mathsf{C}(x,y)$ in 2d (we will comment on the definition in higher dimensions in section~\ref{sec:flatd}) can be defined as
\begin{align}
\label{eq:defC}
\mathsf{C}(x,y)=\left[I_{\mu\alpha'}(x,y)I_{\nu\beta'}(x,y)-g_{\mu\nu}(x)g_{\alpha'\beta'}(y)\right]\langle T^{\mu\nu}(x)T^{\alpha'\beta'}(y) \rangle
\end{align}
where $I_{\mu\alpha'}(x,y)$ is the parallel propagator, $g_{\mu\nu}(x)$ and $g_{\alpha'\beta'}(x)$ are the metric tensor at the two spacetime points. Notice that for Cartesian coordinate in flat spacetime we have $I_{\mu\alpha'}=\delta_{\mu\alpha'}$  and recover the definition (\ref{eq:coord}).\par

Our proposal (\ref{eq:defC}) can be regarded as a covariant point-splitting regularization of the $\rT\overline{\rT}$ operator in curved spacetime. In flat spacetime and Cartesian coordinate it reduces to (\ref{eq:coord}). More importantly, using our proposal it is straightforward to implement the maximal symmetry of spacetime and conservation of the stress energy tensor, as will be demonstrated in the derivations below. In this sense, we believe our proposal is a natural one. On the other hand, there can be other ways to do the point splitting. It is not even necessary to be covariant. While choosing the scheme of point-splitting regularization is a matter of convenience, the final result, which is a statement about the expectation value of the $\rT\overline{\rT}$ operator should be consistent and independent of the choice.

%%%%%%%%%%%%%%%%%%%%%%%%%%%%%%%%%%%%%%%%%%%%%%%
\section{Maximally symmetric bitensors}
\label{sec:maxTensor}
%%%%%%%%%%%%%%%%%%%%%%%%%%%%%%%%%%%%%%%%%%%%%%%
In this section, we discuss the generalizations of the other two ingredients in curved spacetime. To this end, the fact that we are working on the spacetimes with constant curvature is important. The results in this section is valid for general spacetime dimension $d$. Due to the maximal symmetry of the spacetime, the two-point functions of local scalar operators is a function that only depends on the \emph{geodesic distance} between these two points. Namely, we have
\begin{align}
\langle\mathcal{O}_1(x)\mathcal{O}_2(y)\rangle=F(\mathrm{\theta}(x,y))
\end{align}
where $\theta(x,y)$ is the geodesic distance between $x$ and $y$. Similar results hold for two-point functions of operators with spins. In this case, the two-point function is a sum over different tensor structures. The construction of these tensor structures for the two-point functions of maximally symmetric tensors has been studied systematically in \cite{Allen:1985wd}. It is proven that all the tensor structures can be constructed from the vectors $n_{\mu}$ and $m_{\alpha'}$, the parallel propagator $I_{\mu\alpha'}$ and the metric $g_{\mu\nu}$. Here the vectors $n_{\mu}$ and $m_{\alpha'}$ are defined as derivatives of the geodesic distance $\theta(x,y)$ at the two end points
\begin{align}
n_{\mu}(x,y)\equiv \nabla_{\mu}\theta(x,y),\qquad m_{\alpha'}(x,y)=\nabla_{\alpha'}\theta(x,y)
\end{align}
where indices with a prime means we take derivatives at the second position $y$. These two vectors are normalized as $n_{\mu}n^{\mu}=m_{\alpha'}m^{\alpha'}=1$ and are related by the parallel propagator as
\begin{align}
\label{eq:parader}
I_{\mu}^{\phantom{a}\alpha'}m_{\alpha'}+n_{\mu}=0.
\end{align}
Notice that (\ref{eq:parader}) is a generalization of (\ref{eq:xToy}) in flat spacetime. The indices $\mu$ and $\alpha'$ are raised and lowered by the metric at $x$ and $y$ respectively. Using these quantities, the two-point function of the stress energy tensor can be decomposed as
\begin{align}
\label{eq:ansatzTT}
\langle T^{\mu\nu}(x)T^{\alpha'\beta'}(y) \rangle=&\,A_1(\theta)\,n^{\mu}n^{\nu}m^{\alpha'}m^{\beta'}\\\nonumber
&\,+A_2(\theta)\left(I^{\mu\alpha'}n^{\nu}m^{\beta'}+I^{\mu\beta'}n^{\nu}m^{\alpha'}+I^{\nu\alpha'}n^{\mu}m^{\beta'}+I^{\nu\beta'}n^{\mu}m^{\alpha'} \right)\\\nonumber
&\,+A_3(\theta)\left(I^{\mu\alpha'}I^{\nu\beta'}+I^{\mu\beta'}I^{\nu\alpha'}\right)\\\nonumber
&\,+A_4(\theta)\left(n^{\mu}n^{\nu}g^{\alpha'\beta'}+g^{\mu\nu}m^{\alpha'}m^{\beta'} \right)\\\nonumber
&\,+A_5(\theta)\,g^{\mu\nu}g^{\alpha'\beta'}.
\end{align}
where $A_1(\theta),\cdots,A_5(\theta)$ are functions that contain dynamical information of the theory and only depend on the geodesic distance $\theta(x,y)$.
In what follows, we also need the covariant derivatives of the quantities $n_{\mu}, m_{\alpha'}$ and $I_{\mu\alpha'}$. They are given by \cite{Allen:1985wd}
\begin{align}
\label{eq:covD}
\nabla_{\mu}n_{\nu}=&\,\mathcal{A}(\theta)(g_{\mu\nu}-n_{\mu}n_{\nu}),\\\nonumber
\nabla_{\mu}m_{\alpha'}=&\,\mathcal{C}(\theta)(I_{\mu\alpha'}+n_{\mu}m_{\alpha'}),\\\nonumber
\nabla_{\mu}I_{\nu\alpha'}=&\,-(\mathcal{A}(\theta)+\mathcal{C}(\theta))(g_{\mu\nu}m_{\alpha'}+I_{\mu\alpha'}n_{\nu})
\end{align}
where the scalar functions $\mathcal{A}(\theta)$ and $\mathcal{C}(\theta)$ contain information about the spacetime. More explicitly, for different spacetimes, they are
\begin{itemize}
\item Flat spacetime ($\rE_d$ and $\rM_d$)
\[
\mathcal{A}(\theta)=\frac{1}{\theta},\qquad \mathcal{C}(\theta)=-\frac{1}{\theta}
\]
\item Spacetime with positive scalar curvature $d(d-1)/R^2$  ($S^d$ and dS$_d$)
\[
\mathcal{A}(\theta)=\frac{1}{R}\cot\left(\frac{\theta}{R}\right),\qquad \mathcal{C}(\theta)=-\frac{1}{R}\csc\left(\frac{\theta}{R}\right).
\]
\item Spacetime with negative scalar curvature $-d(d-1)/R^2$ ($\rH^d$ and AdS$_d$)
\[
\mathcal{A}(\theta)=\frac{1}{R}\coth\left(\frac{\theta}{R}\right),\qquad \mathcal{C}(\theta)=-\frac{1}{R}\text{csch}\left(\frac{\theta}{R}\right).
\]
\end{itemize}
The bi-vectors $I_{\mu\alpha'}$ are related to the metric by
\begin{align}
g_{\mu\nu}(x)=I_{\mu}^{\phantom{\mu}\alpha'}(x,x')I_{\alpha'\nu}(x',x),\qquad
g_{\alpha'\beta'}(x')=I_{\alpha'}^{\phantom{\alpha}\mu}(x',x)I_{\mu\beta'}(x,x')
\end{align}
The decomposition in (\ref{eq:ansatzTT}) is due to the maximal symmetry of spacetime, this is the generalization of the second point.

\paragraph{Ward identity} Finally, let us consider the implication of conservation of stress-energy tensor
\begin{align}
\nabla_{\mu}T^{\mu\nu}=0.
\end{align}
This leads to Ward identity for the two-point function of the stress-energy tensor. In general, the Ward identity contains some local contact terms. By redefining the operators properly, we can bring the Ward identity to the following form (see for example \cite{Osborn:1999az})
\begin{align}
\label{eq:conserveT}
\nabla_{\mu}\langle T^{\mu\nu}(x)T^{\alpha'\beta'}(y) \rangle=0.
\end{align}
It is understood that our biscalar $\mathsf{C}(x,y)$ is constructed using the operators that satisfies the Ward identity given in (\ref{eq:conserveT}). Acting $\nabla_{\mu}$ on the rhs of (\ref{eq:ansatzTT}) and making use of the relations (\ref{eq:covD}) leads to
\begin{align}
\mathcal{X}\,n^{\nu}m^{\alpha'}m^{\beta'}+\mathcal{Y}(I^{\nu\alpha'}m^{\beta'}+I^{\nu\beta'}m^{\alpha'})+\mathcal{Z}\,n^{\nu}g^{\alpha'\beta'}=0
\end{align}
Since the three tensor structures are independent, this is equivalent to three equations $\mathcal{X}=\mathcal{Y}=\mathcal{Z}=0$ where
\begin{align}
\label{eq:conserved}
\mathcal{X}=&\,A'_1-2A'_2+A'_4+(d-1)\left[\mathcal{A}\,A_1-2(\mathcal{A}+\mathcal{C})\,A_2 \right]+2(\mathcal{A}-\mathcal{C})\,A_2+2\mathcal{C}\,A_4,\\\nonumber
\mathcal{Y}=&\,A'_2-A'_3+d\,\mathcal{A}\,A_2-d(\mathcal{A}+\mathcal{C})\,A_3+\mathcal{C}\,A_4,\\\nonumber
\mathcal{Z}=&\,A'_4+A'_5+(d-1)\mathcal{A}\,A_4+2\mathcal{C}\,A_2-2(\mathcal{A}+\mathcal{C})\,A_3.
\end{align}
Here $d$ is the dimension of spacetime and $A'_i\equiv A'_i(\theta)=\frac{ d A_i(\theta)}{d\theta}$.
The invariant biscalar $\mathsf{C}(x,y)$ can also be written in terms of $A_i(\theta)$. Using the definition (\ref{eq:defC}) and (\ref{eq:ansatzTT}), it is straightforward to find that
\begin{align}
\label{eq:CAii}
\mathsf{C}(x,y)=2(1-d)A_2+d(d-1)A_3+2(1-d)A_4+d(1-d)A_5
\end{align}

%%%%%%%%%%%%%%%%%%%%%%%%%%%%%%%%%%%%%%%%%%%%%%%%
\section{The invariant biscalar in 2d}
\label{sec:invariantC2d}
%%%%%%%%%%%%%%%%%%%%%%%%%%%%%%%%%%%%%%%%%%%%%%%%
In this section, we focus on two dimensional spacetime and see the implication of the spacetime symmetry and conservation of stress energy tensor on the biscalar $\mathsf{C}(x,y)$. We will comment on higher dimensions in section~\ref{sec:flatd}. Taking $d=2$ in (\ref{eq:CAii}), we have
\begin{align}
\label{eq:Cxy2}
\mathsf{C}(x,y)=-2(A_2-A_3+A_4+A_5)
\end{align}
The constraints from conservation of the stress tensor simplify to
\begin{align}
\label{eq:2dconserved}
\mathcal{X}=&\,A'_1-2A'_2+A'_4+\left[\mathcal{A}\,A_1-2(\mathcal{A}+\mathcal{C})\,A_2 \right]+2(\mathcal{A}-\mathcal{C})\,A_2+2\mathcal{C}\,A_4=0,\\\nonumber
\mathcal{Y}=&\,A'_2-A'_3+2\,\mathcal{A}\,A_2-2(\mathcal{A}+\mathcal{C})\,A_3+\mathcal{C}\,A_4=0,\\\nonumber
\mathcal{Z}=&\,A'_4+A'_5+\mathcal{A}\,A_4+2\mathcal{C}\,A_2-2(\mathcal{A}+\mathcal{C})\,A_3=0.
\end{align}
Similar to (\ref{eq:partialC}), we consider $\partial_{\mu}\mathsf{C}(x,y)$ where we take derivative with respect to $x$. From (\ref{eq:Cxy2}), we have
\begin{align}
\label{eq:dC}
\partial_{\mu}\mathsf{C}(x,y)=-2(A'_2-A'_3+A'_4+A'_5)n_{\mu}
\end{align}
Now it is interesting to see that using the constraint $\mathcal{Y}+\mathcal{Z}=0$ we have
\begin{align}
A'_2-A'_3+A'_4+A'_5+(\mathcal{A}+\mathcal{C})(2A_2-4A_3+A_4)=0
\end{align}
This enables us to get ride of the derivatives $A'_i$ in (\ref{eq:dC}) completely and write
\begin{align}
\label{eq:curveDC}
\partial_{\mu}\mathsf{C}(x,y)=2(\mathcal{A}+\mathcal{C})(2A_2-4A_3+A_4)n_{\mu}
\end{align}
The combination $2A_2-4A_3+A_4$ can be written as another projection of the two-point function of stress-energy tensor
\begin{align}
2A_2-4A_3+A_4=-\left(I_{\mu\alpha'}I_{\nu\beta'}-n_{\mu}n_{\nu}g_{\alpha'\beta'}\right)\langle T^{\mu\nu}(x)T^{\alpha'\beta'}(y)\rangle
\end{align}
Defining another biscalar
\begin{align}
\mathsf{D}(x,y)=\left(I_{\mu\alpha'}I_{\nu\beta'}-n_{\mu}n_{\nu}g_{\alpha'\beta'}\right)\langle T^{\mu\nu}(x)T^{\alpha'\beta'}(y)\rangle,
\end{align}
we arrive at the following equation
\begin{align}
\label{eq:main}
\partial_{\mu}\mathsf{C}(x,y)+2n_{\mu}(\mathcal{A}+\mathcal{C})\mathsf{D}(x,y)=0.
\end{align}
This is our main equation for the biscalar $\mathsf{C}(x,y)$ which is the curved spacetime generalization of (\ref{eq:partialC}). It is a result of the symmetry of the spacetime and conservation of the stress-energy tensor and is valid for all constant curvature spacetime in 2d. Several comments are in order.\par

In flat spacetime, we have $\mathcal{A}(\theta)+\mathcal{C}(\theta)=\theta^{-1}-\theta^{-1}=0$ and the second term in (\ref{eq:main}) vanishes. Therefore we simply have $\partial_{\mu}\mathsf{C}(x,y)=0$. This is an alternative proof that $\mathsf{C}(x,y)$ is a constant in flat spacetime. Our proof here does not rely on any specific coordinate system.

In the large-$c$ limit, due to the factorization, we have
\begin{align}
\mathsf{D}(x,y)=&\,\left(I_{\mu\alpha'}I_{\nu\beta'}-n_{\mu}n_{\nu}g_{\alpha'\beta'}\right)\langle T^{\mu\nu}(x)T^{\alpha'\beta'}(y)\rangle\\\nonumber
\sim&\,\left(I_{\mu\alpha'}I_{\nu\beta'}-n_{\mu}n_{\nu}g_{\alpha'\beta'}\right)\langle T^{\mu\nu}\rangle\langle T^{\alpha'\beta'}\rangle\\\nonumber
\sim&\left(I_{\mu\alpha'}I_{\nu\beta'}-n_{\mu}n_{\nu}g_{\alpha'\beta'}\right)g^{\mu\nu}g^{\alpha'\beta'}=0
\end{align}
where in the third line we used the fact that $\langle T^{\mu\nu}\rangle\propto g^{\mu\nu}$ for maximally symmetric spacetime. So the second term of (\ref{eq:main}) again vanishes. This shows that $\mathsf{C}(x,y)$ is indeed a constant in the large-$c$ limit which is consistent with large-$c$ factorization.\par

For curved spacetime $\mathcal{A}+\mathcal{D}\ne 0$ and finite $c$ it is clear that $\mathsf{D}(x,y)$ is not vanishing in general. Therefore for these cases, $\mathsf{C}(x,y)$ is no longer a constant. Nevertheless, we can use it to obtain an expression for the expectation value of the $\langle\rT\overline{\rT}\rangle$ operator.

%%%%%%%%%%%%%%%%%%%%%%%%%%%%%%%%%%%%%%%%%%%%%%%%%%%%%%%%%%%%%%
\section{Expectation value of $\rT\overline{\rT}$ operator}
\label{sec:expectV}
%%%%%%%%%%%%%%%%%%%%%%%%%%%%%%%%%%%%%%%%%%%%%%%%%%%%%%%%%%%%%%
In this section, we give an expression for the expectation value of $\langle\rT\overline{\rT}\rangle$ operator using our result (\ref{eq:main}). Let us first comment on the coinciding limit $y\to x$. In this limit, the two-point function is governed by the structure of OPE. Since the structure of OPE is a local property of the given quantum field theory, it does not depend on the curvature of the spacetime. Therefore, the coinciding limit analysis is exactly the same as in the flat spacetime. The $\rT\overline{\rT}$ operator is as well-defined as in the flat spacetime case up to total derivatives. Since we are considering homogenous space, these total derivatives will not affect the expectation values as in the flat spacetime. We therefore have
\begin{align}
\mathsf{C}(x,x)=\lim_{y\to x}\mathsf{C}(x,y)=\langle\rT\overline{\rT}\rangle
\end{align}
for constant curvature spacetimes.\par

For later convenience, let us define the following quantity
\begin{align}
\mathsf{\Delta}(x,y)=\mathsf{C}(x,y)-\mathsf{D}(x,y)
\end{align}
It can be written as the following projection of the two-point function of the stress energy tensor
\begin{align}
\mathsf{\Delta}(x,y)=&\,-(g_{\mu\nu}-n_\mu n_{\nu})\,g_{\alpha'\beta'}\langle T^{\mu\nu}(x)T^{\alpha'\beta'}(y)\rangle\\\nonumber
=&\,-(g_{\mu\nu}-n_{\mu}n_{\nu})\langle T^{\mu\nu}(x)\Theta(y)\rangle
\end{align}
where we have defined $\Theta(y)=g_{\alpha'\beta'}(y)T^{\alpha'\beta'}(y)$. We can rewrite (\ref{eq:main}) as
\begin{align}
\partial_{\mu}\mathsf{C}(x,y)+2(\mathcal{A}+\mathcal{C})\mathsf{C}(x,y)n_{\mu}=2(\mathcal{A}+\mathcal{C})\mathsf{\Delta}(x,y)n_{\mu}
\end{align}
This equation can be further rewritten as
\begin{align}
\label{eq:intEq}
\partial_{\mu}\left[f(\theta)\mathsf{C}(x,y)\right]=g(\theta)\mathsf{\Delta}(x,y)n_{\mu}
\end{align}
where
\begin{align}
f(\theta)=&\,\alpha\,\exp\left[2\int[\mathcal{A}(\theta)+\mathcal{C}(\theta)]d\theta\right],\\\nonumber
g(\theta)=&\,2[\mathcal{A}(\theta)+\mathcal{C}(\theta)]f(\theta)=df(\theta)/d\theta.
\end{align}
Here $\alpha$ is a multiplicative constant. For different spacetimes, the functions $f(\theta)$ and $g(\theta)$ are given by
\begin{itemize}
\item Flat spacetime
\[
f(\theta)=\alpha,\qquad g(\theta)=0
\]
\item Positive curvature
\[
f(\theta)=\alpha \left[\cos\left(\frac{\theta}{2R}\right)\right]^4,\qquad g(\theta)=-\frac{2\alpha}{R}\left[\cos\left(\frac{\theta}{2R}\right)\right]^3\left[\sin\left(\frac{\theta}{2R}\right)\right]
\]
\item Negative curvature
\[
f(\theta)=\alpha \left[\cosh\left(\frac{\theta}{2R}\right)\right]^4,\qquad g(\theta)=\frac{2\alpha}{R}\left[\cosh\left(\frac{\theta}{2R}\right)\right]^3\left[\sinh\left(\frac{\theta}{2R}\right)\right]
\]
\end{itemize}
Multiplying both sides of (\ref{eq:intEq}) by $n^{\mu}$, we have
\begin{align}
n^{\mu}\partial_{\mu}\left[f(\theta)\mathsf{C}(x,y)\right]=g(\theta)\mathsf{\Delta}(x,y)
\end{align}
Using the fact that $n^{\mu}\nabla_{\mu}=d/d\theta$
\begin{align}
\frac{d}{d\theta}\left[f(\theta)\mathsf{C}(x,y)\right]=g(\theta)\mathsf{\Delta}(x,y)
\end{align}
Integrating both sides for $x$ from $y$ to some $x=y_c$ and denoting $\theta_c=\theta(y_c,y)$, we have
\begin{align}
f(\theta_c)\mathsf{C}(y_c,y)-f(0)\mathsf{C}(y,y)=\int_0^{\theta_c}g(\theta)\mathsf{\Delta}(x,y)d\theta
\end{align}
Notice that we have $\mathsf{C}(y,y)=\langle\rT\overline{\rT}\rangle$ and $f(0)=\alpha$. Rewriting the above equation, we find
\begin{align}
\label{eq:TTbEq}
\langle\rT\overline{\rT}\rangle=\alpha^{-1}f(\theta_c)\mathsf{C}(y_c,y)-\alpha^{-1}\int_0^{\theta_c}g(\theta)\mathsf{\Delta}(x,y)d\theta
\end{align}
Our equation (\ref{eq:TTbEq}) is a generalization of Zamolodchikov's result to spacetimes with non-zero curvature. Plugging in the functions $f(\theta)$ and $g(\theta)$ for different spacetime, we arrive at
\begin{itemize}
\item Flat spacetime
\begin{align}
\langle\rT\overline{\rT}\rangle=\mathsf{C}(y,y_c)
\end{align}
\item Positive curvature
\begin{align}
\label{eq:positive}
\langle\rT\overline{\rT}\rangle=\left(\cos\frac{\theta_c}{2R}\right)^4\mathsf{C}(y,y_c)
+\frac{2}{R}\int_0^{\theta_c}\left(\sin\frac{\theta}{2R}\right)\left(\cos\frac{\theta}{2R}\right)^3\mathsf{\Delta}(x,y)d\theta
\end{align}
\item Negative curvature
\begin{align}
\label{eq:negative1}
\langle\rT\overline{\rT}\rangle=\left(\cosh\frac{\theta_c}{2R}\right)^4\mathsf{C}(y,y_c)
-\frac{2}{R}\int_0^{\theta_c}\left(\sinh\frac{\theta}{2R}\right)\left(\cosh\frac{\theta}{2R}\right)^3\mathsf{\Delta}(x,y)d\theta
\end{align}
\end{itemize}
We see that, by taking the $R\to\infty$ limit for the spacetimes with non-zero curvature, we indeed recover the flat spacetime result at the leading order.\par

Since the spacetime is homogeneous, the lhs of (\ref{eq:positive}) and (\ref{eq:negative1}) are constant. Therefore the rhs should also be constant. This implies that the rhs is in fact independent of $\theta_c$ and we can choose the value of $\theta_c$ at our convenience.\par

For the positive curvature spacetime, we can choose $\theta_c=\pi R$. For a 2-sphere, this happens when $y$ and $y_c$ are at antipodal points. Then the first term vanishes and we have
\begin{align}
\label{eq:antipodal}
\langle\rT\overline{\rT}\rangle=
\frac{2}{R}\int_0^{\pi R}\left(\sin\frac{\theta}{2R}\right)\left(\cos\frac{\theta}{2R}\right)^3\mathsf{\Delta}(x,y)d\theta
\end{align}
Using $\langle T^{\mu\nu}(x)\rangle=C\,g^{\mu\nu}(x)$, the above equation can be written as
\begin{align}
\langle\rT\overline{\rT}\rangle=-2C^2+\frac{2}{R}\int_0^{\pi R}\left(\sin\frac{\theta}{2R}\right)\left(\cos\frac{\theta}{2R}\right)^3\mathsf{\Delta}_{\text{con}}(x,y)d\theta
\end{align}
where we have used the fact that $g(\theta)=df(\theta)/d\theta$ and $\mathsf{\Delta}_{\text{conn}}(x,y)$ is the connected part of $\mathsf{\Delta}(x,y)$ defined by
\begin{align}
\label{eq:connectDelta}
\mathsf{\Delta}_{\text{con}}(x,y)=-(g_{\mu\nu}-n_{\mu}n_{\nu})\left[\langle T^{\mu\nu}(x)\Theta(y)\rangle-\langle T^{\mu\nu}(x)\rangle\langle\Theta(y)\rangle \right].
\end{align}
This is our final expression for $\langle\rT\overline{\rT}\rangle$ in positive curvature spacetime. The first term on the rhs is the factorized result at large $c$ or in flat spacetime. The second term gives the deviation from the factorized result which is due to spacetime curvature.\par

For the negative curvature spacetime, since they are non-compact, we can choose $\theta_c\to\infty$. This means we take $y$ and $y_c$ to be infinitely far away from each other. In this case, the first term factorizes due to clustering property of correlation functions
\begin{align}
\label{eq:negative}
\langle\rT\overline{\rT}\rangle=\lim_{\theta_c\to\infty}\left(\cosh\frac{\theta_c}{2R}\right)^4\mathsf{C}(y,y_c)
-\frac{2}{R}\int_0^{\infty}\left(\sinh\frac{\theta}{2R}\right)\left(\cosh\frac{\theta}{2R}\right)^3\mathsf{\Delta}(x,y)d\theta
\end{align}
This is also the choice for the flat spacetime where one derives the factorization formula. The first term in (\ref{eq:negative}) is divergent in the limit $\theta_c\to\infty$. This is fine since the second term is also divergent and the divergence should cancel to give a finite answer. Using $\langle T^{\mu\nu}(x)\rangle=C\,g^{\mu\nu}(x)$, we have
\begin{align}
\label{eq:thetaInft}
\lim_{\theta_c\to\infty}\mathsf{C}(y_c,y)=-2C^2.
\end{align}
Using the fact that the disconnected part of $\mathsf{\Delta}(x,y)$ is also $-2C^2$, we can rewrite the integrand of (\ref{eq:negative}) as
\begin{align}
\label{eq:rewriteInt}
\int_0^{\theta_c}\rg(\theta)(\Delta_{\text{con}}-2C^2)d\theta=\int_0^{\theta_c}\rg(\theta)\mathsf{\Delta}_{\text{con}}(x,y)d\theta-2C^2(f(\theta_c)-1)
\end{align}
Combining (\ref{eq:rewriteInt}) with (\ref{eq:thetaInft}), we see that in the limit $\theta_c\to\infty$, we have
\begin{align}
\langle\rT\overline{\rT}\rangle=&\,f(\theta_c)(-2C^2)-\int_0^{\theta_c}\rg(\theta)\mathsf{\Delta}_{\text{con}}(x,y)d\theta+2C^2(f(\theta_c)-1)\\\nonumber
=&\,-2C^2-\int_0^{\theta_c}\rg(\theta)\mathsf{\Delta}_{\text{con}}(x,y)d\theta
\end{align}
We find that $\langle\rT\overline{\rT}\rangle$ in the two cases can be written collectively as
\begin{align}
\label{eq:TTcol}
\langle\rT\overline{\rT}\rangle=\langle\rT\rangle\langle\overline{\rT}\rangle-\int_0^{\theta_{\text{max}}}\rg(\theta)\mathsf{\Delta}_{\text{con}}(x,y)d\theta
\end{align}
where the first term is the factorized result and $\rg(\theta)$ is $g(\theta)$ by setting $\alpha=1$. The second term characterize the deviation from large-$c$ or flat spacetime where $\theta_{\text{max}}$ is the maximal possible value for the geodesic distance in the spacetime. For positive curvature space, it the case when $x$ and $y$ are at antipodal points and $\theta_{\text{max}}=\pi R$; for negative curvature space, this is infinity.
To compute $\langle\rT\overline{\rT}\rangle$ at finite $c$, we need to know $\mathsf{\Delta}(x,y)$ and then integrate along a geodesic with a specific weight function $g(\theta)$. Since $\mathsf{\Delta}(x,y)$ is a specific projection of the two-point function of the stress-energy tensor, we see that in general the expectation value $\langle\rT\overline{\rT}\rangle$ will depend on the information of \emph{two-point function} of the stress energy tensor in curved spacetime. This is in contrast to the situation of flat spacetime where we can write $\langle\rT\overline{\rT}\rangle$ simply in terms of \emph{one-point function} of stress-energy tensor, which is the first term of the rhs in (\ref{eq:TTcol}). One possible implication of this result is that the $\rT\overline{\rT}$ deformation in curved spacetime is no longer solvable in the sense of flat spacetime. However, the finite $c$ correction is characterized by a specific projection $\mathsf{\Delta}_{\text{con}}$ which is relatively simple. Therefore the corresponding $\rT\overline{\rT}$ deformation might still exhibit some simplicity.

It might happen in some special situations we have a better control on $\mathsf{\Delta}(x,y)$, then we can do better in computing the expectation value $\langle\rT\overline{\rT}\rangle$. As a simple example, we consider the expectation value of $\rT\overline{\rT}$ operator for CFTs on curved spacetime. In this case, we have a further simplification\footnote{One can define the stress-energy tensor properly to get rid of trace anomaly and write the trace equation in this form. For more details, see \cite{Osborn:1999az}.}
\begin{align}
\mathsf{\Delta}(x,y)=-(g_{\mu\nu}-n_{\mu}n_{\nu})\langle T^{\mu\nu}(x)\Theta(y)\rangle=0
\end{align}
where we have used the traceless condition for CFT. In this case, we simply have
\begin{align}
\label{eq:diffF}
\partial_{\mu}\left[f(\theta)\mathsf{C}(x,y)\right]=0.
\end{align}
This leads to
\begin{align}
\mathsf{C}(x,y)=\frac{\alpha}{\cos^4(\theta/(2R))}
\end{align}
for positive curvature spacetime. Here $\alpha$ is an integration constant. Since there is no reason to expect a divergence at $\theta=\pi R$, we must put $\alpha=0$ for this case and we have
\begin{align}
\langle\rT\overline{\rT}\rangle=0.
\end{align}
For negative curvature spacetime, we have
\begin{align}
\mathsf{C}(x,y)=\frac{\alpha}{\cosh^4(\theta/(2R))}
\end{align}
and we have
\begin{align}
\langle\rT\overline{\rT}\rangle=\alpha
\end{align}
In the large-$c$ limit, $\alpha$ is given by $-2C^2$.

%%%%%%%%%%%%%%%%%%%%%%%%%%%%%%%%%%%%%%%%%%%%%%%%%%%%
\section{Examples in explicit coordinate system}
\label{sec:example}
%%%%%%%%%%%%%%%%%%%%%%%%%%%%%%%%%%%%%%%%%%%%%%%%%%%%
Our discussions in the previous sections are general and does not refer to explicit coordinate systems. In this section, we write down our main results in some coordinate systems. This will enable us to write down some of the results more explicitly. Since geodesic distance $\theta(x,y)$ plays an essential role in our construction, we first discuss how to compute it in maximally symmetric spacetimes.

All the maximally symmetric spacetimes can be embedded in a higher dimensional flat spacetime by imposing quadratic constraints. Denoting the embedding coordinate by $X^A$, $A=0,1,\cdots,d$. The maximally symmetric spaces are defined by imposing the following constraints
\begin{align}
\label{eq:embed}
\eta_{AB}X^AX^B=a^2.
\end{align}
Depending on the choice of signature for $\eta_{AB}$ and whether $a$ is real or purely imaginary, we obtain different spacetimes. One can choose proper intrinsic coordinate $x^{\mu}$ such that $X^A(x)$ satisfies (\ref{eq:embed}). The geodesic distance $\theta(x,y)$ between two points $x^{\mu}$ and $y^{\mu}$ are given by the following formula
\begin{align}
\label{eq:def-dis}
\cos\left(\frac{\theta(x,y)}{a}\right)=\frac{\eta_{AB}X^A(x)X^B(y)}{a^2}
\end{align}
This is nothing but a direct generalization of the well-known formula in Euclidean space
\begin{align}
\label{eq:geodis}
\cos\varphi=\frac{\mathbf{x}\cdot\mathbf{y}}{|\mathbf{x}||\mathbf{y}|}
\end{align}
where $\varphi$ is the angle between two vectors $\mathbf{x}$ and $\mathbf{y}$. From the constraint (\ref{eq:embed}) we see that the norm of the `vector' is simply $a$. As a simple example, let us consider the 2-sphere $S^2$. In this case, $a=R$ and the metric $\eta_{AB}$ is given by
\begin{align}
\eta_{AB}=\text{diag}(+1,+1,+1)
\end{align}
and the intrinsic coordinate $(\phi,\varphi)$ is related to the embedding coordinate by
\begin{align}
X^0(\phi,\varphi)=R\sin\phi\sin\varphi,\qquad X^1(\phi,\varphi)=R\sin\phi\cos\varphi,\qquad X^2(\phi,\varphi)=R\cos\phi
\end{align}
According to (\ref{eq:geodis}), we have
\begin{align}
\label{eq:distance}
\cos\left(\frac{\theta(x,y)}{R}\right)=\frac{1}{R^2}\sum_{i=0}^2X^i(\phi,\varphi)X^i(\phi',\varphi')
=\cos\phi\cos\phi'+\sin\phi\sin\phi'\cos(\varphi-\varphi')
\end{align}
Therefore for sphere we have
\begin{align}
\label{eq:sphregeodesic}
\theta(x,y)=R\,\arccos\left( \cos\phi\cos\phi'+\sin\phi\sin\phi'\cos(\varphi-\varphi')\right)
\end{align}
The choices of $a$ and $\eta_{AB}$ for different spacetimes are given by
\begin{itemize}
\item Two-sphere ${S}^2$
\begin{align}
\eta_{AB}=\text{diag}(+1,+1,+1),\qquad a=R
\end{align}
\item de Sitter space dS$_2$
\begin{align}
\eta_{AB}=\text{diag}(-1,+1,+1),\qquad a=R
\end{align}
\item Hyperbolic space $\rH^2$
\begin{align}
\eta_{AB}=\text{diag}(-1,+1,+1),\qquad a=iR
\end{align}
\item Anti-de Sitter space AdS$_2$
\begin{align}
\eta_{AB}=\text{diag}(-1,+1,-1),\qquad a=iR
\end{align}
\end{itemize}
Using the formula (\ref{eq:distance}), it is straightforward to compute the geodesic distance $\theta(x,y)$. One comment is that for the cases where there is no geodesic between $x$ and $y$, we should take (\ref{eq:def-dis}) as the definition of $\theta$. Taking derivatives with respect to $x$ and $y$ give $n_{\mu}$ and $m_{\alpha'}$. To compute the parallel propagator, we can simply use (\ref{eq:covD}) and rewrite
\begin{align}
I_{\mu\alpha'}=\mathcal{C}(\theta)^{-1}\partial_{\mu}m_{\alpha'}-n_{\mu}m_{\alpha'}
=\mathcal{C}(\theta)^{-1}\partial_{\mu}\partial_{\alpha'}\theta-\partial_{\mu}\theta\partial_{\alpha'}\theta.
\end{align}
With expressions of $I_{\mu\alpha'}$, $n_{\mu}$ and $g_{\mu\nu}$, we can write down the biscalars $\mathsf{C}(x,y)$ and $\mathsf{\Delta}(x,y)$ explicitly.

\subsection{Flat spacetime}
Before going to the curved spacetime, we revisit the flat spacetime in polar coordinate system and compute $\mathsf{C}(x,y)$ explicitly. The main point is that the parallel propagator $I_{\mu\alpha'}$ is non-trivial in the polar coordinate system. This coordinate system is useful for studying $\rT\overline{\rT}$ deformations on a disc. In the polar coordinate $x^{\mu}=(r,\varphi)$ the metric is
\begin{align}
ds^2=dr^2+r^2 d\varphi^2.
\end{align}
The geodesic distance between the points $x^{\mu}=(r,\varphi)$ and $y^{\mu}=(r',\varphi')$ is given by
\begin{align}
\theta(x,y)=\sqrt{r^2+{r'}^2-2rr'\cos(\varphi-\varphi')}
\end{align}
It is straightforward to compute the vectors $n_{\mu}=(n_r,n_{\varphi})$ and $m_{\alpha'}=(m_r,m_{\varphi})$ with
\begin{align}
&n_r=\frac{r-r'\cos(\varphi-\varphi')}{\theta(x,y)},\qquad n_{\varphi}=\frac{r r' \sin(\varphi-\varphi')}{\theta(x,y)},\\\nonumber
&m_r=\frac{r'-r\cos(\varphi-\varphi')}{\theta(x,y)},\qquad m_{\varphi}=-\frac{r r' \sin(\varphi-\varphi')}{\theta(x,y)}.
\end{align}
It is easy to verify that these vectors are normalized $n_\mu n^\mu=m_{\alpha'}m^{\alpha'}=1$. Using these, we find
\begin{align}
I_{rr}=&\,\cos(\varphi-\varphi'),&\quad I_{r\varphi}=&\,r'\sin(\varphi-\varphi'),\\\nonumber
I_{\varphi r}=&\,-r\sin(\varphi-\varphi'),&
\quad I_{\varphi\varphi}=&\,r r'\cos(\varphi-\varphi').
\end{align}
We can now write down $\mathsf{C}(x,y)$ explicitly. Without loss of generality, we can put $y$ at the origin $(r',\varphi')=(0,0)$. To simplify the expression, we introduce the shorthand notation
\begin{align}
\rT^{\mu\nu;\alpha'\beta'}=\langle T^{\mu\nu}(x)T^{\alpha'\beta'}(0)\rangle
\end{align}
Then the biscalar is given by
\begin{align}
\label{eq:CCpolar}
\mathsf{C}(x,0)=&\,-\sin^2(\varphi)\,\rT^{rr;rr}
-r\sin(2\varphi)\,\rT^{r\varphi;rr}
-r^2\cos^2(\varphi)\,\rT^{\varphi\varphi;rr}
\end{align}
We can compare this to the Cartesian coordinate expression
\begin{align}
\mathsf{C}(x,0)=-\rT^{11;22}-\rT^{22;11}+2\rT^{12;12}
\end{align}
and see that the later does not involve any non-trivial coefficient in front of $\rT^{\mu\nu;\alpha\beta}$. However, in polar coordinate, in order $\mathsf{C}$ to be constant, we need non-trivial coefficients in front of the various components of the two-point function.

\subsection{The sphere}
In this subsection, we consider the spacetimes with positive curvature. We focus on the 2-sphere with radius $R$. We do the computation in two coordinate systems.

\paragraph{Coordinate system I} In the spherical coordinate system, the metric is given by
\begin{align}
ds^2=R^2d\phi^2+R^2\sin^2\phi\,d\varphi^2
\end{align}
The points on the sphere are parameterized by $x=(\phi,\varphi)$. The geodesic distance between two points $x=(\phi,\varphi)$ and $y=(\phi',\varphi')$ has been derived in (\ref{eq:sphregeodesic}) and takes the following form
\begin{align}
\label{eq:disSphere}
\theta(x,y)=R\,\arccos\left[\cos\phi\cos\phi'+\sin\phi\sin\phi'\cos(\varphi-\varphi')\right]
\end{align}
We can easily compute $n_{\mu}=(n_{\phi},n_{\varphi})$ and $m_{\alpha'}=(m_{\phi},m_{\varphi})$ where
\begin{align}
n_{\phi}=&\,\frac{R(\sin\phi\cos\phi'-\cos\phi\sin\phi'\cos(\varphi-\varphi'))}{\sqrt{1-[\cos\phi\cos\phi'+\sin\phi\sin\phi'\cos(\varphi-\varphi')]^2}},\\\nonumber
n_{\varphi}=&\,\frac{R\sin\phi\sin\phi'\sin(\varphi-\varphi')}{\sqrt{1-[\cos\phi\cos\phi'+\sin\phi\sin\phi'\cos(\varphi-\varphi')]^2}}
\end{align}
and
\begin{align}
m_{\phi}=&\,\frac{R(\cos\phi\sin\phi'-\sin\phi\cos\phi'\cos(\varphi-\varphi'))}{\sqrt{1-[\cos\phi\cos\phi'+\sin\phi\sin\phi'\cos(\varphi-\varphi')]^2}},\\\nonumber
m_{\varphi}=&\,-\frac{R\sin\phi\sin\phi'\sin(\varphi-\varphi')}{\sqrt{1-[\cos\phi\cos\phi'+\sin\phi\sin\phi'\cos(\varphi-\varphi')]^2}}.
\end{align}
The components of the parallel propagator are given by
\begin{align*}
I_{\phi\phi}=&\,R^2\frac{(1+\cos\phi\cos\phi')\cos(\varphi-\varphi')+\sin\phi\sin\phi'}{1+\cos\phi\cos\phi'+\cos(\varphi-\varphi')\sin\phi\sin\phi'},\\\nonumber
I_{\varphi\phi}=&\,-R^2\frac{(\cos\phi+\cos\phi')\sin(\varphi-\varphi')\sin\phi}{1+\cos\phi\cos\phi'+\cos(\varphi-\varphi')\sin\phi\sin\phi'},\\\nonumber
I_{\phi\varphi}=&\,R^2\frac{(\cos\phi+\cos\phi')\sin(\varphi-\varphi')\sin\phi'}{1+\cos\phi\cos\phi'+\cos(\varphi-\varphi')\sin\phi\sin\phi'},\\\nonumber
I_{\varphi\varphi}=&\,R^2\frac{\left[(1+\cos\phi\cos\phi')\cos(\varphi-\varphi')+\sin\phi\sin\phi'\right]\sin\phi\sin\phi'}
{1+\cos\phi\cos\phi'+\cos(\varphi-\varphi')\sin\phi\sin\phi'}.
\end{align*}
Without loss of generality, we can put $y$ at the origin with $(\phi',\varphi')=(0,\varphi')$. Then the quantities simplifies considerably. We have
\begin{align}
\theta(x,0)=R\phi,\qquad d\theta=Rd\phi
\end{align}
and
\begin{align}
n_{\mu}=&\,(n_{\phi},n_{\varphi})=(R{\sin\phi}/{|\sin\phi|},0),\\\nonumber
m_{\alpha'}=&\,(m_{\phi},m_{\varphi})=(-R\cos(\varphi-\varphi')\sin\phi/|\sin\phi|,0)
\end{align}
The parallel propagators simplify to
\begin{align}
I_{\phi\phi}=R^2\cos(\varphi-\varphi'),\qquad I_{\varphi\phi}=-R^2\sin\phi\sin(\varphi-\varphi'),\qquad I_{\phi\varphi}=I_{\varphi\varphi}=0.
\end{align}
We therefore have
\begin{align}
\label{eq:Csphere}
\mathsf{C}(x,0)=&\,-R^4\sin^2(\varphi-\varphi')\rT^{\phi\phi;\phi\phi}-R^4\sin^2\phi\cos^2(\varphi-\varphi')
\rT^{\varphi\varphi;\phi\phi}\\\nonumber
&\,-R^4\sin\phi\sin(2(\varphi-\varphi'))\rT^{\phi\varphi;\phi\phi}
\end{align}
and
\begin{align}
\mathsf{\Delta}(x,0)=-R^4\sin^2\phi\,\rT^{\varphi\varphi;\phi\phi}
\end{align}
The functions $f(\theta)$ and $g(\theta)$ become (taking $\alpha=1$)
\begin{align}
f(\theta)=\left(\cos\frac{\phi}{2}\right)^4,\qquad g(\theta)=-\frac{2}{R}\sin\frac{\phi}{2}\left(\cos\frac{\phi}{2}\right)^4
\end{align}
Using the result (\ref{eq:antipodal}) and noticing that $\theta(x,0)=R\phi$ in our case, we have
\begin{align}
\langle\rT\overline{\rT}\rangle=\langle\rT\rangle\langle\overline{\rT}\rangle+R^4\int_0^{\pi}(\sin\phi)^3\left(\cos\frac{\phi}{2}\right)^2\langle T^{\varphi\varphi}(x)T^{\phi\phi}(0)\rangle_{\text{con}}\, d\phi
\end{align}
This is a rather compact expression for the expectation value of $\rT\overline{\rT}$ operator.

\paragraph{Coordinate system II}
We consider another coordinate system for the sphere $S^2$ which can be generalized to negative curvature space. Let us define
\begin{align}
X_1=R\frac{2\rho\cos\varphi}{1+\rho^2},\qquad X_2=R\frac{2\rho\sin\varphi}{1+\rho^2},\qquad X_3=R\frac{1-\rho^2}{1+\rho^2}
\end{align}
The geodesic distance, by putting $y$ to the origin $(\rho',\varphi')=(0,\varphi')$ is
\begin{align}
\theta(x,0)=R\arccos\left(\frac{1-\rho^2}{1+\rho^2}\right)=2R\arctan\rho,\qquad d\theta=\frac{2R}{1+\rho^2}d\rho.
\end{align}
The geodesic distance is independent of $\varphi,\varphi'$. This implies that $n_{\varphi}(x,0)$ is vanishing. The vectors $n_{\mu}=(n_\rho,n_{\varphi})$ and $m_{\alpha}=(m_{\rho},m_{\varphi})$ are given by
\begin{align}
n_{\rho}=\frac{2R}{1+\rho^2},\qquad n_{\varphi}=0,\qquad m_{\rho}=-2R\cos\Delta\varphi,\qquad m_{\varphi}=0.
\end{align}
where $\Delta\varphi=\varphi-\varphi'$. The components of the parallel propagator are
\begin{align}
I_{\rho\rho}=\frac{4R^2\cos\Delta\varphi}{1+\rho^2},\qquad I_{\rho\varphi}=0,\qquad I_{\varphi\rho}=-\frac{4R^2\rho\sin\Delta\varphi}{1+\rho^2},\qquad I_{\varphi\varphi}=0.
\end{align}
We find the following result for $\mathsf{C}(x,0)$
\begin{align}
\mathsf{C}(x,0)=&\,-\frac{16R^4\sin^2\Delta\varphi}{(1+\rho^2)^2}\,\rT^{\rho\rho;\rho\rho}
-\frac{16R^4\rho^2\cos^2\Delta\varphi}{(1+\rho^2)^2}\,\rT^{\varphi\varphi;\rho\rho}
-\frac{16R^4\rho\sin2\Delta\varphi}{(1+\rho^2)^2}\,\rT^{\rho\varphi;\rho\rho}
\end{align}
and
\begin{align}
\mathsf{\Delta}(\rho)=-\frac{16R^4\rho^2}{(1+\rho^2)^2}\,\rT^{\varphi\varphi;\rho\rho}
\end{align}
The functions $f(\theta)$ and $g(\theta)$ in this coordinate become (setting $\alpha=1$)
\begin{align}
f(\theta)=\frac{1}{(1+\rho^2)^2},\qquad g(\theta)=-\frac{2}{R}\frac{\rho}{(1+\rho^2)^2}
\end{align}
The limit $\theta_c=\pi R$ is given by $\rho\to\infty$. We can similarly write down the expectation value of the $\rT\overline{\rT}$ operator
\begin{align}
\langle\rT\overline{\rT}\rangle=\langle\rT\rangle\langle\overline{\rT}\rangle+64R^4\int_0^{\infty}\frac{\rho^3\,d\rho}{(1+\rho^2)^5}\langle T^{\varphi\varphi}(x)T^{\rho\rho}(0)\rangle_{\text{con}}
\end{align}

\subsection{The Poincar\'e disc}
In this subsection, we consider spacetimes with constant negative curvature. We consider the Poincar\'e disc, which is similar to the second coordinate system of the sphere in the previous subsection. The embedding coordinates are given by
\begin{align}
X_1=R\frac{2\rho\cos\varphi}{1-\rho^2},\qquad X_2=R\frac{2\rho\sin\varphi}{1-\rho^2},\qquad X_3=R\frac{1+\rho^2}{1-\rho^2}.
\end{align}
The intrinsic metric is given by
\begin{align}
ds^2=\frac{4R^2}{(1-\rho^2)^2}\left(d\rho^2+\rho^2d\varphi^2\right)
\end{align}
As before, we put the point $y$ at $(\rho',\varphi')=(0,\varphi')$. The geodesic distance is given by
\begin{align}
\theta(x,0)=R\,\text{arccosh}\left(\frac{1+\rho^2}{1-\rho^2}\right)=2R\,\text{arctanh}(\rho),\qquad d\theta=\frac{2R}{1-\rho^2}d\rho
\end{align}
The vectors $n_{\mu}=(n_{\rho},n_{\varphi})$ and $m_{\alpha}=(m_{\rho},m_{\varphi})$ are given by
\begin{align}
n_{\rho}=\frac{2R}{1-\rho^2},\qquad n_{\varphi}=0,\qquad m_{\rho}=2R\cos\Delta\varphi,\qquad m_{\varphi}=0.
\end{align}
The parallel propagators are given by
\begin{align}
I_{\rho\rho}=-\frac{4R^2\cos\Delta\varphi}{1-\rho^2},\qquad I_{\varphi\rho}=\frac{4R^2\rho\sin\Delta\varphi}{1-\rho^2},\qquad I_{\rho\varphi}=I_{\varphi\Delta\varphi}=0.
\end{align}
The quantity $\mathsf{C}(x,0)$ and $\mathsf{\Delta}(x,0)$ are given by
\begin{align}
\label{eq:CDexplicit}
\mathsf{C}(x,0)=&\,-\frac{16R^4\sin^2\Delta\varphi}{(1-\rho^2)^2}\rT^{\rho\rho;\rho\rho}
-\frac{16R^4\rho^2\cos^2\Delta\varphi}{(1-\rho^2)^2}\rT^{\varphi\varphi;\rho\rho}
-\frac{16R^4\rho\sin(2\Delta\varphi)}{(1-\rho^2)^2}\rT^{\rho\varphi;\rho\rho}
\end{align}
and
\begin{align}
\mathsf{\Delta}(x,0)=-\frac{16R^4\rho^2}{(1-\rho^2)^2}\rT^{\varphi\varphi;\rho\rho}.
\end{align}
The functions $f(\theta)$ and $g(\theta)$ are given by
\begin{align}
f(\theta)=\frac{1}{(1-\rho^2)^2},\qquad g(\theta)=\frac{2}{R}\frac{\rho}{(1-\rho^2)^2}
\end{align}
The limit $\theta_c\to\infty$ corresponds to $\rho\to 1$. In this case, we have
\begin{align}
\langle\rT\overline{\rT}\rangle=\langle\rT\rangle\langle\overline{\rT}\rangle+64R^4\int_0^1\frac{\rho^3 d\rho}{(1-\rho^2)^5}\langle T^{\varphi\varphi}(x)T^{\rho\rho}(0)\rangle_{\text{con}}
\end{align}

%%%%%%%%%%%%%%%%%%%%%%%%%%%%%%%%%%%%%%%%%%%%%%%%%%%%%%%%%%%%%%%%%%%%%%%%
\section{Comments on other dimensions}
\label{sec:flatd}
%%%%%%%%%%%%%%%%%%%%%%%%%%%%%%%%%%%%%%%%%%%%%%%%%%%%%%%%%%%%%%%%%%%%%%%%
In this section, we make some comments on similar analysis in other dimensions. We want to highlight some simplifications that only occurs in 2d. For simplicity, we focus on the flat spacetime in $d$ dimensions. In this case we have $\mathcal{A}+\mathcal{C}=0$ but $d\ne2$. The constraints become
\begin{align}
\label{eq:conserved}
\mathcal{X}=&\,A'_1-2A'_2+A'_4+\left[(d-1)A_1+4A_2+2A_4\right]\theta^{-1}=0,\\\nonumber
\mathcal{Y}=&\,A'_2-A'_3+\left[d\,A_2-A_4\right]\theta^{-1}=0,\\\nonumber
\mathcal{Z}=&\,A'_4+A'_5+\left[(d-1)\,A_4-2A_2\right]\theta^{-1}=0.
\end{align}
In $d$ dimensions, let us search for a slightly more general definition for the biscalar
\begin{align}
\mathsf{C}_d(x,y)=\left[I_{\mu\alpha'}I_{\nu\beta'}+a_d\,g_{\mu\nu}g_{\alpha'\beta'}\right]\langle T^{\mu\nu}(x)T^{\alpha'\beta'}(y)\rangle
\end{align}
where $a_d$ is some constant depend on dimension $d$. We can write this quantity in terms of $A_i(\theta)$ as
\begin{align}
\label{eq:Cd}
\mathsf{C}_d(x,y)=&\,(a_d+1)A_1-2(2a_d+d+1)A_2+d(2a_d+d+1)A_3\\\nonumber
&\,+(2a_d d+2)A_4+d(a_d d+1)A_5.
\end{align}
As before we want to study the quantity $\partial_{\mu}\mathsf{C}_d(x,y)$ using the constraints in (\ref{eq:conserved}). The nice thing happens in 2d is that we can get rid of the derivatives $A'_i(\theta)$ completely by using the constraints and write $\partial_{\mu}\mathsf{C}$ in terms of only $A_i(\theta)$. We will show that this is impossible in higher dimensions for any choice of $a_d$.\par

Noticing that $A'_1$ only appears in $\mathcal{X}$, $A'_3$ only appears in $\mathcal{Y}$ and $A'_5$ appears only in $\mathcal{Z}$ and the form of (\ref{eq:Cd}), we should take the following combination of the constraint
\begin{align}
(a_d+1)\mathcal{X}+d(2a_d+d+1)\mathcal{Y}+d(a_d d+1)\mathcal{Z}=0
\end{align}
and get rid of $A'_1,A'_3$ and $A'_5$ in $\nabla_{\mu}\mathsf{C}_d(x,y)$. This leads to
\begin{align}
\label{eq:dmuC}
\nabla_{\mu}\mathsf{C}_d(x,y)=&\,(d-1)(2a_d+d)A'_2\,n_{\mu}-(d-1)((d-1)a_d+1)A'_4\,n_{\mu}\\\nonumber
&\,-\theta^{-1}(d-1)(a_d+1)A_1\,n_{\mu}\\\nonumber
&\,+\theta^{-1}(d-1)(d^2+2d+4+4(d+1)a_d)A_2\,n_{\mu}\\\nonumber
&\,+\theta^{-1}(d-1)(2(d+1)+(d^2+2)a_d)A_4\,n_{\mu}.
\end{align}
We see that we still have two derivative terms $A'_2(\theta)$ and $A'_4(\theta)$. There are two cases where the above expression simplifies further. One is $a_d=1/(1-d)$. In this case, $A'_4$ is also absent and we have
\begin{align}
\label{eq:bis1}
\nabla_{\mu}\mathsf{C}^{(1)}_d(x,y)=&\,(d^2-d-2)A'_2\,n_{\mu}-\theta^{-1}(d-2)A_1\,n_{\mu}\\\nonumber
&\,+\theta^{-1}(d-2)(d^2+3d+4)A_2\,n_{\mu}-\theta^{-1}(d^2-4)A_4\,n_{\mu}
\end{align}
If we want to further get rid of $A'_2(\theta)$ from the above equation, we can only choose $d=2$ or $d=-1$. Since we have $d\ge 1$, so $d=2$ is the only possible choice.\par

The other choice of $a_d$ which simplifies(\ref{eq:dmuC}) is $a_d=-d/2$, in this case $A'_2$ is absent and we have
\begin{align}
\label{eq:bis2}
\nabla_{\mu}\mathsf{C}^{(2)}_d(x,y)=&\,\frac{1}{2}(d^2-1)(d-2)A'_4\, n_{\mu}\\\nonumber
&\,+\frac{\theta^{-1}}{2}(d-1)(d-2)A_1\,n_{\mu}-\theta^{-1}(d^2-4)(d-1)A_2\,n_{\mu}\\\nonumber
&\,+\frac{\theta^{-1}}{2}(d-1)(d^3-2d-4)A_4\,n_{\mu}.
\end{align}
If we want to further eliminate the term with $A'_4(\theta)$ in the above equation, we can take $d=1,2$. Two comments are in order.
Firstly, it is interesting to notice that the biscalar $\mathsf{C}_d^{(1)}(x,y)$ defined in (\ref{eq:bis1}) is identical to the one motivated from gravity \cite{Taylor:2018xcy,Hartman:2018tkw,Caputa:2019pam}. Whether this is a coincidence or there are some reasons behind is unclear to us at the moment. Secondly, for $\mathsf{C}^{(2)}_d$, there are two special values for $d$. One is $d=2$ which reproduces the 2d result. The second is $d=1$. In this case, we also have $\partial_{\mu}\mathsf{C}^{(2)}_1(x,y)=0$. This seems to indicate that one can define $\rT\overline{\rT}$ deformation for 1D QFT, or quantum mechanics. This is somewhat trivial but might be of interest in the context of $NAdS_2/NCFT_1$ duality.\par

The fact that we can no longer get rid of all the $A'_i$ in $\nabla_{\mu}\mathsf{C}^{(1)}(x,y)$ shows that there's a qualitative difference between $d=2$ and $d>2$. In fact, one should not be surprised by this since in higher dimensions the coinciding limit of $\mathsf{C}^{(1)}_d(x,y)$ is not well-defined. It can be seen already by considering $\mathsf{C}^{(1)}(x,y)$ for CFT where
\begin{align}
\mathsf{C}^{(1)}_d(x,y)=\langle T(x)\bar{T}(y)\rangle\sim\frac{d-2}{(x-y)^2}.
\end{align}
For $d\ne 2$, the coinciding limit $\mathsf{C}^{(1)}_d(x,x)$ is divergent and one needs to perform regularization. More discussions on these points can be found in \cite{Taylor:2018xcy,Hartman:2018tkw,Caputa:2019pam}.

%%%%%%%%%%%%%%%%%%%%%%%%%%%%%%%%%%%%%%%%%%%%%%%%%%
\section{Conclusions and outlook}
\label{sec:conclude}
%%%%%%%%%%%%%%%%%%%%%%%%%%%%%%%%%%%%%%%%%%%%%%%%%%
We considered the expectation value of the $\rT\overline{\rT}$ operator in spacetimes with constant curvature. We defined an invaraint biscalar using two-point function of the stress-energy tensor and the parallel propagator. Using the symmetry of spacetime and the conservation of the stress-energy tensor, we derived a differential equation for the biscalar. From this equation, we show that in flat spacetime and in the large-$c$ limit, the biscalar is a constant, which leads to the factorization formula. In spacetimes with non-zero curvature and at finite $c$, the biscalar is not a constant. We can nevertheless write down an expression for $\langle\rT\overline{\rT}\rangle$ which shows that in general it also depends on the information of two-point functions of the stress-energy tensor.\par

On the one hand, the results in this paper is in a sense negative, showing that the expectation value of $\langle\rT\overline{\rT}\rangle$ is more complicated than in the flat spacetime case and does not factorize at finite $c$. On the other hand, we have an explicit expression for the deviation from large $c$ which only depends on $\mathsf{\Delta}_{\text{con}}$. It is interesting to study the $\rT\overline{\rT}$ deformation of QFT in curved spacetime based on our result. It is expected that the deformation will be more complicated than the flat spacetime case, but it might as well be that it is still simple enough to be studied analytically to some extent. For example, can one gain some better understanding for $\mathsf{\Delta}_{\text{con}}$ in some special cases and find the perturbative $1/c$ corrections for the factorized result ?\par

As a byproduct of our analysis in curved spacetime, we have a better understanding why the factorization formula works in flat spacetime and large $c$ limit. In the flat spacetime, our result gives a good starting point to analyse the $\rT\overline{\rT}$ deformation in other coordinates such as polar coordinate which are more suitable for other geometries. It will be interesting to revisit the $\rT\overline{\rT}$ deformation for these cases.\par

Finally, it should be possible to apply our method to other irrelevant deformations which are triggered by higher dimensional operators. For example, the integrable deformations triggered by higher conserved charges in integrable quantum field theories.

%%%%%%%%%%%%%%%%%%%%%%%%%%%%%%%%%%%%%%%%%%%%%%%%%%
\section*{Acknowldegement}
\label{sec:thanks}
%%%%%%%%%%%%%%%%%%%%%%%%%%%%%%%%%%%%%%%%%%%%%%%%%%
It is our pleasure to thank Ofer Aharony, Alex Belin, John Cardy, Pawel Caputa, Shouvik Datta, David Kutasov, Amit Giveon, Kostas Siampos, Amit Sever, Sasha Zhiboedov and Yang Zhou for helpful discussions and correspondences. We also thank Shouvik Datta for collaboration at the initial stage of this project. We thank the warm hospitality of ITP-CAS and University of Milano-Bicocca where part of the work is done.

\appendix

%%%%%%%%%%%%%%%%%%%%%%%%%%%%%%%%%%%
\section{Some useful formula}
%%%%%%%%%%%%%%%%%%%%%%%%%%%%%%%%%%%
In this appendix, we collect some formulas that are useful for the computations in the main text. The following formulas states that $I_{\mu\alpha}$, $n_\mu$ and $m_\alpha$ are covariant constant along the geodesic
\begin{align}
n^{\mu}\nabla_{\mu}n_{\nu}=0,\qquad n^{\mu}\nabla_{\mu}m_{\alpha}=0,\qquad n^{\mu}\nabla_{\mu}I_{\nu\alpha}=0.
\end{align}
For contracting indices, we have
\begin{align}
\nabla_{\mu}n^{\mu}=(d-1)\mathcal{A},\qquad \nabla^{\mu}I_{\mu\alpha}=-(d-1)(\mathcal{A}+\mathcal{C})m_{\alpha}
\end{align}
The two-point functions can be projected to different tensor structures. To this end, we consider the following tensors
\begin{align}
&\rE^{(1)}_{\mu\nu\alpha'\beta'}=n_{\mu}n_{\nu}m_{\alpha'}m_{\beta'},&\\\nonumber
&\rE^{(2)}_{\mu\nu\alpha'\beta'}=I_{\mu\alpha'}I_{\nu\beta'},\\\nonumber
&\rE^{(3)}_{\mu\nu\alpha'\beta'}=I_{\mu\alpha'}n_{\nu}m_{\beta'},\\\nonumber
&\rE^{(4)}_{\mu\nu\alpha'\beta'}=g_{\mu\nu}m_{\alpha'}m_{\beta'},\\\nonumber
&\rE^{(5)}_{\mu\nu\alpha'\beta'}=g_{\mu\nu}g_{\alpha'\beta'}.
\end{align}
The contractions of these tensors with the two-point functions of the stress energy tensor are given by
\begin{align}
B_1\equiv\rE^{(1)}_{\mu\nu\alpha'\beta'}\langle T^{\mu\nu}(x)T^{\alpha'\beta'}(y)\rangle=&\,A_1-4A_2+2A_3+2A_4+A_5,\\\nonumber
B_2\equiv\rE^{(2)}_{\mu\nu\alpha'\beta'}\langle T^{\mu\nu}(x)T^{\alpha'\beta'}(y)\rangle=&\,A_1-2(d+1)A_2+d(d+1)A_3+2A_4+d A_5,\\\nonumber
B_3\equiv\rE^{(3)}_{\mu\nu\alpha'\beta'}\langle T^{\mu\nu}(x)T^{\alpha'\beta'}(y)\rangle=&\,-A_1+(d+3)A_2-(d+1)A_3-2A_4p-A_5,\\\nonumber
B_4\equiv\rE^{(4)}_{\mu\nu\alpha'\beta'}\langle T^{\mu\nu}(x)T^{\alpha'\beta'}(y)\rangle=&\,A_1-4A_2+2A_3+(d+1)A_4+dA_5,\\\nonumber
B_5\equiv\rE^{(5)}_{\mu\nu\alpha'\beta'}\langle T^{\mu\nu}(x)T^{\alpha'\beta'}(y)\rangle=&\,A_1-4A_2+2dA_3+2dA_4+d^2 A_5.
\end{align}
For general $d$, the five $B_i$ are linear independent. For $d=2$, we have
\begin{align}
B_2+2B_3+2B_4-B_5=0
\end{align}

\providecommand{\href}[2]{#2}\begingroup\raggedright\endgroup

%\bibliographystyle{JHEP}
%\bibliography{yunfeng}

\end{document}